\documentclass[10pt]{IEEEtran}

\usepackage{graphicx}   
 \usepackage{graphicx}
 \graphicspath{{eps/}} \DeclareGraphicsExtensions{.eps}

\usepackage{url}        

\usepackage{amsmath}    
\usepackage[all]{xy}
\usepackage{calc}
\usepackage{epic,eepic}
\usepackage{amssymb,eufrak}
\usepackage{xcolor}
\usepackage{accents}
\usepackage{marvosym}
\usepackage{comment}
\usepackage{mathbbol}
\usepackage{bbm}
\usepackage{mathrsfs}

\newtheorem{theo}{Theorem}
\newtheorem{prop}[theo]{Proposition}

\newtheorem{coro}[theo]{Corollary}

\newtheorem{exam}[theo]{Example}

\newcommand{\N}{{\cal H}}

\newcommand{\C}{\mathscr{C}}
\renewcommand{\L}{\mathscr{L}}
\newcommand{\gf}{\mbox{\rm\sc gf}}
\newcommand{\pr}{\mbox{\rm Pr}}
\newcommand{\card}{\mbox{\rm card}}

\newcommand{\npIn}{n' \in {\cal H}(m)\setminus\{n\}}
\newcommand{\mIn}{m \in {\cal H}(n)}
\newcommand{\mpIn}{m' \in {\cal H}(n)\setminus\{m\}}
\newcommand{\sumover}{\stackrel{(a_{n'})_{n'\in\N(m)}}{\in\L(m\mid a_n = a)}}
\newcommand{\dist}{\mbox{\rm\small dist}}

\newcommand{\AI}{\mbox{\rm\sc ai}}
\newcommand{\COT}{\mbox{\rm\sc cot}}

\renewcommand{\max}{\mathop{\mbox{\rm max}}}
\renewcommand{\min}{\mathop{\mbox{\rm min}}}
\newcommand{\argmax}{\mathop{\mbox{\rm argmax}}}
\newcommand{\argmin}{\mathop{\mbox{\rm argmin}}}

\newcommand{\msazero}{\mbox{Min-Sum}_0}
\newcommand{\msastar}{\mbox{Min-Sum}_{\star}}

\newcommand{\topspace}[1][4.25mm]{\vbox{\hbox{\vspace{#1}}}}

\newenvironment{mylist}[1][$\bullet$]{\begin{list}{#1}{\itemindent .5\parindent \leftmargin 0mm }}{\end{list}}

\begin{document}
%
\title{Min-Max decoding for non binary LDPC codes}
%
%
\author{Valentin~Savin, CEA-LETI, MINATEC, Grenoble, France, valentin.savin@cea.fr
\thanks{Part of this work was performed in the frame of the %
ICTNEWCOM++ project which is a partly EU funded Network of %
Excellence of the 7th European Framework Program.}
}
\maketitle

\begin{abstract}
Iterative decoding of non-binary LDPC codes is currently performed using either the Sum-Product or the Min-Sum algorithms or slightly
different versions of them. In this paper, several low-complexity quasi-optimal iterative algorithms are proposed for decoding non-binary
codes. The Min-Max algorithm is one of them and it has the benefit of two possible LLR domain implementations: a {\em standard
implementation}, whose complexity scales as the square of the Galois field's cardinality and a reduced complexity implementation called
{\em selective implementation}, which makes the Min-Max decoding very attractive for practical purposes.
\end{abstract}


%
\IEEEpeerreviewmaketitle

\section{Introduction}
%
%
%
%

Although already proposed by Gallager in his PhD thesis \cite{gall_phd}, \cite{gall-monograph} the use of non binary LDPC codes is still
very limited today, mainly because of the decoding complexity of these codes. The optimal iterative decoding is performed by the
Sum-Product algorithm \cite{Wiberg} at the price of an increased complexity, computation instability, and dependence on thermal noise
estimation errors. The Min-Sum algorithm \cite{Wiberg} performs a sub-optimal iterative decoding, less complex than the Sum-Product
decoding, and independent of thermal noise estimation errors. The sub-optimality of the Min-Sum decoding comes from the overestimation of
extrinsic messages computed within the check-node processing.

The Sum-Product algorithm can be efficiently implemented in the probability domain using binary Fourier transforms
\cite{barnault_fast_2003} and its complexity is dominated by ${\cal O}(q\log_2q)$ sum and product operations for each check node
processing, where $q$ is the cardinality of the Galois field of the non-binary LDPC code. The Min-Sum decoding can be implemented either in
the log-probability domain or in the log-likelihood ratio (LLR) domain and its complexity is dominated by ${\cal O}(q^2)$ sum operations
for each check node processing. In the LLR domain, a reduced selective implementation of the Min-Sum decoding, called Extended Min-Sum, was
proposed in \cite{declercq_extended_2005}, \cite{declercq_decoding_2007}. Here ``selective'' means that the check-node processing uses the
incoming messages concerning only a part of the Galois field elements. Non binary LDPC codes were also investigated in \cite{davey1998ldp},
\cite{hu2003ctg}, \cite{wymeersch2004ldd}.

In this paper we propose several new algorithms for de\-co\-ding non-binary LDPC codes, one of which is called the Min-Max algorithm. They
are all independent of thermal noise estimation errors and perform quasi-optimal decoding -- meaning that they present a very small
performance loss with respect to the optimal iterative decoding (Sum-Product). We also propose two implementations of the Min-Max
algorithm, both in the LLR domain, so that the decoding is computationally stable: a {\it ``standard implementation''} whose complexity
scales as the square of the Galois field's cardinality and a reduced complexity implementation called {\it ``selective implementation''}.
That makes the Min-Max decoding very attractive for practical purposes.

The paper is organized as follows. In the next section we briefly review several realizations of the Min-Sum algorithm for non binary LDPC
codes. It is intended to keep the paper self-contained but also to justify some of our choices regarding the new decoding algorithms
introduced in section \ref{sec:Min-Norm}. The implementation of the Min-Max decoder is discussed in section \ref{sec:implementations}.
Section \ref{sec:simulations} presents simulation results and section \ref{sec:conclusions} concludes this paper.

The following notations will be used throughout the paper.

\smallskip\noindent{\bf Notations related to the Galois field:}
\begin{list}{$\bullet$}{\itemindent 0mm \leftmargin\parindent}
\item $\gf(q) = \{0,1,\dots,q-1\}$, the {\it Galois field with $q$ elements}, where $q$ is a power of a prime number. Its elements will be called
{\it symbols}, in order to be distinguished from ordinary integers.
\item $a, s, x$ will be used to denote $\gf(q)$-symbols.
\item $\mathbb{a, s, x}$ will be used to denote vectors of $\gf(q)$-symbols.
For instance, $\mathbb{a} = (a_1, \dots, a_I) \in \gf(q)^{I}$, etc.
\end{list}

\smallskip\noindent {\bf Notations related to LDPC codes:}
\begin{list}{$\bullet$}{\itemindent 0mm \leftmargin\parindent}
\item $H\in\mathbf{M}_{M,N}(\gf(q))$, the {\it $q$-ary check matrix} of the code.
\item $\C$, set of codewords of the LDPC code.
\item $\C_n(a)$, set of codewords with the $n^{\mbox{\scriptsize th}}$ coordinate equal to $a$; for given  $1\leq n \leq N$ and $a\in\gf(q)$.
\item $\mathbb{x} = (x_1,x_2,\dots,x_N)$ a {\it $q$-ary codeword} transmitted over the channel.
\end{list}

The Tanner graph associated with an LDPC code consists of $N$ {\it variable nodes} and $M$ {\it check nodes} representing the $N$ columns
and the $M$ lines of the matrix $H$. A variable node and a check node are connected by an edge if the corresponding element of matrix $H$
is not zero. Each edge of the graph is labeled by the corresponding non zero element of $H$.

\smallskip\noindent {\bf Notations related to the Tanner graph:}
\begin{list}{$\bullet$}{\itemindent 0mm \leftmargin\parindent}
\item ${\cal H}$, the Tanner graph of the code.
\item $n \in \{ 1, 2, \dots, N \}$ a {\it variable node} of ${\cal H}$.
\item $m \in \{ 1, 2, \dots, M \}$ a {\it check node} of ${\cal H}$.
\item ${\cal H}(n)$, set of neighbor check nodes of the variable node $n$.
\item ${\cal H}(m)$,\!\hfill set\!\hfill of\!\hfill neighbor\!\hfill variable\!\hfill nodes\!\hfill of\!\hfill the\!\hfill check\!\hfill node\!\hfill $m$.
\item $\L(m)$,  set of {\it local configurations} verifying the check node $m$; {\em i.e.} the set of sequences of $\gf(q)$-symbols\break
$\mathbb{a} = (a_n)_{n\in {\cal H}(m)}$ , verifying the linear constraint:
$$ \sum_{n\in {\cal H}(m)} h_{m,n}a_n = 0$$
\item $\L(m \mid a_n = a)$,  set of local configurations  $\mathbb{a}$ verifying $m$, such that $a_n = a$; for given $n\in {\cal H}(m)$ and
$a\in\gf(q)$.
\end{list}

An iterative decoding algorithm consists of an initialization step followed by an iterative exchange of messages between variable and check
nodes connected in the Tanner graph. 

\smallskip\noindent {\bf Notations related to an iterative decoding algorithm:}
\begin{list}{$\bullet$}{\itemindent 0mm \leftmargin\parindent}
\item $\gamma_n(a)$, the {\it a priori information of the variable node $n$
concerning the symbol $a$}.
\item $\tilde{\gamma}_n(a)$, the {\it a posteriori information of the variable node $n$
concerning the symbol $a$}.
\item $\alpha_{m,n}(a)$, the {\it message from the check node $m$ to the variable node $n$ concerning the symbol
$a$.}
\item $\beta_{m,n}(a)$, the {\it message from the variable node $n$ to the check node $m$ concerning the symbol
$a$.}
\end{list}

\section{Realizations of the Min-Sum decoding\\ for non binary LDPC codes}

\subsection{Min-Sum decoding}\label{subsec:min_sum}
The Min-Sum decoding is generally implemented in the log-probability domain and it performs the following operations:

\smallskip\noindent {\bf Initialization} \\
{\it $\bullet$ A priori information}:
 $\gamma_n(a) = -\ln\left(\pr( x_n = a \mid \mbox{channel} )\right)$
{\it $\bullet$ Variable node messages}:
 $\alpha_{m,n}(a) = \gamma_n(a)$

\smallskip\noindent{\bf Iterations}\\
{\it $\bullet$ Check node processing}
$$\beta_{m,n}(a) = \displaystyle\min_{\sumover} \left(\sum_{\npIn} \alpha_{m,n'}(a_{n'})\right)$$
{\it $\bullet$ Variable node processing}
$$\alpha_{m,n}(a) = \displaystyle \gamma_n(a) + \sum_{\mpIn} \beta_{m',n}(a)$$
{\it $\bullet$ A posteriori information}
$$\tilde{\gamma}_n(a) = \displaystyle \gamma_n(a) + \sum_{\mIn} \beta_{m,n}(a)$$

\smallskip For practical purposes, messages $\alpha_{m,n}(a)$ and $\beta_{m,n}(a)$ should be normalized in order to avoid computational instability
(otherwise they could ``escape'' to infinity). The check node processing, which dominates the decoding complexity, can be implemented using
a {\it forward -- backward} computation method.

Assuming that the Tanner graph is cycle free, the a posteriori information converges after finitely many iterations to \cite{Wiberg}:
$$\tilde{\gamma}_n(a) = \displaystyle\min_{\mathbb{a}\in\C_n(a)}
   \left(\sum_{k = 1}^{N}\gamma_k(a_k)\right)$$
Moreover, if $s_n = \displaystyle \argmin_{a\in\gf(q)} \left(\tilde{\gamma}_n(a)\right)$ is the most likely symbol according to the a
posteriori information computed above, then the sequence $\mathbb{s} = (s_1, s_2, \dots, s_N)$ is a codeword (this is no longer true if the
Tanner graph contains cycles) and it can be easily verified that $\mathbb{s} = \argmax_{\mathbb{a}\in\C}\pr\left(\mathbb{x} =
\mathbb{a}\mid \mbox{channel}\right)$.

\medskip 
\noindent Thus, under the cycle free assumption the Min-Sum decoding always outputs a codeword and the above equality looks rather like a
maximum likelihood decoding, which is in contrast with the sub-optimality property. In \cite{Wiberg}, the author concludes that the Min-Sum
decoding is optimal in terms of {\it block error probability}, but our explanation is quite different. What really happens is that
probabilities from the above equality do not take into account the dependence relations between codeword's symbols (or equivalently between
graph's variable nodes). Taking into account only the channel observation and no prior information about dependence relations between
codeword's symbols, we have $\pr\left(\mathbb{x} = \mathbb{a}\mid \mbox{channel}\right) = \displaystyle\prod_{n=1}^{N} \pr\left(x_n =
a_n\mid \mbox{channel}\right)$, therefore even a sequence $\mathbb{x}$ which is not a codeword has a non zero probability. In some sense,
the Min-Sum decoding converges to a {\it maximum likelihood decoding, distorted by dependence relations between codeword's symbols}.

\subsection{Equivalent iterative decoders}
The term of {\it equivalent (iterative) decoders} will be employed several times through this paper. We begin this section by providing its
rigorous definition. Consider the a posteriori information available at a variable node $n$ after the $l^{\mbox{\scriptsize th}}$ decoding
iteration: it defines an order between the symbols of the Galois field, starting with the most likely symbol and ending with the least
likely one. Note that the most likely symbol may correspond to the minimum or to the maximum of the a posteriori information, depending on
the decoding algorithm. We call this order the {\em a posteriori order} of variable node $n$ at iteration $l$.

We say that {\em two decoders are equivalent} if, for each variable node $n$, they both induce the same a posteriori order at any iteration
$l$. In particular, assuming that the hard decoding corresponds to the most likely symbol, both decoders output the same sequence of
$\gf(q)$-symbols.

Two iterative decoders, which are equivalent to the Min-Sum decoder are presented below.

\subsubsection{$\msazero$ decoding} The $\msazero$ decoding performs the following operations.

\smallskip\noindent {\bf Initialization} \\
{\it $\bullet$ A priori information}\footnote{We assume that all codewords are sent equally likely.}
$$\gamma_n(a) = \ln\left({\pr( x_n = 0 \mid \mbox{channel} )} /
                               {\pr( x_n = a \mid \mbox{channel} )}\right)$$
{\it $\bullet$ Variable node messages}:
 $\alpha_{m,n}(a) = \gamma_n(a)$

\smallskip\noindent{\bf Iterations}\\
{\it $\bullet$ Check node processing}:
 same as for Min-Sum (\ref{subsec:min_sum})

\smallskip\noindent{\it $\bullet$ Variable node processing}
$$\begin{array}{rrl}
\alpha'_{m,n}(a) & = & \displaystyle \gamma_n(a) + \sum_{\mpIn} \beta_{m',n}(a) \\
\alpha_{m,n}(a) & = & \alpha'_{m,n}(a) - \alpha'_{m,n}(0)
\end{array}$$
 {\it $\bullet$ A posteriori information}: same as for Min-Sum (\ref{subsec:min_sum})

We note that the exchanged messages represent log-likelihood ratios with respect to a fixed symbol (here, the symbol $0\in\gf(q)$ is used,
but obviously other symbols may be considered). Its main advantage with respect to the classical Min-Sum decoding is that it is
computationally stable, so that there is no need to normalize the exchanged messages. This algorithm is also known as the Extended Min-Sum
algorithm \cite{declercq_extended_2005}, \cite{declercq_decoding_2007}.

\subsubsection{$\msastar$ decoding} The $\msastar$ decoding performs the following operations.

\smallskip\noindent {\bf Initialization} \\
{\it $\bullet$ A priori information}
$$\gamma_n(a) = \ln\left({\pr( x_n = s_n \mid \mbox{channel} )} /
                               {\pr( x_n = a \mid \mbox{channel} )}\right)$$
where $s_n$ is the most likely symbol for $x_n$.

\smallskip\noindent {\it $\bullet$ Variable node messages}:
 $\alpha_{m,n}(a) = \gamma_n(a)$

\smallskip\noindent{\bf Iterations}\\
{\it $\bullet$ Check node processing}: same as for Min-Sum (\ref{subsec:min_sum})
%

\smallskip\noindent{\it $\bullet$ Variable node processing}
$$\begin{array}{rrl}
\alpha'_{m,n}(a) & = & \displaystyle \gamma_n(a) + \sum_{\mpIn} \beta_{m',n}(a) \\
\alpha'_{m,n}    & = & \displaystyle \min_{a\in\gf(q)}\alpha'_{m,n}(a) \\
 \alpha_{m,n}(a) & = & \alpha'_{m,n}(a) - \alpha'_{m,n}
\end{array}$$
 {\it $\bullet$ A posteriori information}: same as for Min-Sum (\ref{subsec:min_sum})

As the $\msazero$ decoding, the $\msastar$ decoding also performs in the LLR domain and is computationally stable. The main difference is
that the exchanged messages represent log-likelihood ratios with respect to the most likely symbol, which may vary from a variable node to
another, or within a fixed variable node, from an iteration to the other.

\begin{theo}
The Min-Sum, $\msazero$ and $\msastar$ decoders are equivalent.
\end{theo}

It follows that the $\msastar$ decoding does not present any practical interest: the equivalent $\msazero$ decoding is already
computationally stable and less complex (it does not require the minimum computation in the variable node processing step). However, the
$\msastar$ motivates the decoding algorithms that will be introduced in the following section. The fundamental observation is that messages
concerning most likely symbols are always equal to zero, and messages concerning the other symbols are positive. Therefore, exchanged
messages can be seen as metrics indicating how far is a given symbol from the most likely one. This will be developed in the next section.

\section{Min-Norm decoding for non binary LDPC codes}
\label{sec:Min-Norm}

According to the discussion in the above section, we interpret variable node messages as metrics indicating the distance between a given
symbol and the most likely one. Consider a check node $m$ and a variable node $n\in\N(m)$. Using the received extrinsic information, {\em
i.e.} metrics received from variable nodes $n'\in\N(m)\setminus\{n\}$, the check node $m$ has to evaluate:
\begin{mylist}
\item the most likely symbol for variable node $n$,
\item how far the other symbols are from the most likely one.
\end{mylist}

To simplify notation, we set $\N(m) = \{n, n_1, \dots,n_d\}$ and let $s_{i}$ be the most likely symbols corresponding to variable node
$n_i$, $i=1,\dots,d$. Since the linear constraint corresponding to the check node $m$ has to be satisfied, the most likely symbol for the
variable node $n$ is the unique symbol $s\in\gf(q)$, such that:
$$(s,s_1, \dots, s_d) \in \L(m)$$
On the other hand, each symbol $a\in\gf(q)$ corresponds to a set of $d$-tuples $(a_1,\cdots, a_d)$, such that $(a, a_1, \dots, a_d) \in
\L(m)$:
$$\L_a(m) = \{(a_1, \dots,a_d) \mid (a, a_1, \dots, a_d) \in \L(m)\}$$

Thus, identifying $s\equiv(s_1,\dots,s_d)$ and $a\equiv \L_a(m)$, the distance between the most likely symbol $s$ and the symbol $a$ can be
evaluated as the distance from the sequence $(s_1,\dots,s_d)$ to the set $\L_a(m)$. As usual, we consider that the distance from a point to
a set is equal to the minimum distance between the given point and the points of the given set. In addition, we have to specify a rule that
computes the distance between two sequences $(s_1,\dots,s_d)$ and $(a_1,\dots,a_d)$, taking into account the received ``marginal
distances'' between $s_i$ and $a_i$'s. This can be done by using the distance associated with one of the $p$-norms ($p\geq 1$), or the
infinity (also called maximum) norm on an appropriate multidimensional vector space. Precisely :
$$ \begin{array}{c@{\,}l}
  \beta_{m,n}(a) & = \dist\left( (s_1,\dots,s_d),\,{\cal L}_a(m)\right)\\
           &  = \displaystyle\min_{(a_1,\dots,a_d)\in {\cal L}_a(m)} \dist\left( (s_1,\dots,s_d),\,(a_1,\dots,a_d)\right)\topspace[4.5mm]\\
           &  = \displaystyle\min_{(a_1,\dots,a_d)\in {\cal L}_a(m)}
             \parallel\dist(s_1,a_1),\dots,\dist(s_d,a_d)\parallel_p\topspace[4.5mm]\\
           &  = \displaystyle\min_{(a_1,\dots,a_d)\in {\cal L}_a(m)}
             \parallel\alpha_1(a_1),\dots,\alpha_d(a_d)\parallel_p\topspace[4.5mm]
  \end{array}$$

Note that for $p = 1$ we obtain the $\msastar$ decoding from the above section. Taking into account that:
$$ \parallel\ \parallel_{1} \ \geq \ \parallel\ \parallel_2 \ \geq \  \cdots  \ \geq \  \parallel\ \parallel_\infty $$
it follows that using any of the $\parallel\ \parallel_{p}$ or $\parallel\ \parallel_\infty$ norms would reduce the value of check node
messages, which is de\-si\-rable, since these messages are actually overestimated by the $\msastar$ algorithm. One could think that the
infinity norm would excessively reduce check node messages values, but this is not true. In fact, the distance between $\parallel\
\parallel_{1}$ and the infinity norm is less than twice the distance between $\parallel\ \parallel_{1}$ and the Euclidian ($p = 2$) norm.
In practice, it turns out that the infinity norm induces a more accurate computation of check node messages than $\parallel\
\parallel_1$.

We focus now on the Euclidean and the infinity norms. The derived decoding algorithms are called {\em Euclidean de\-co\-ding}, respectively
{\em Min-Max decoding}. They perform the same initialization step, variable nodes processing, and a posteriori information update as the
$\msastar$ decoding; therefore only the check node processing step will be described below.

\subsection{Euclidean decoding}

\noindent
{\it $\bullet$ Check nodes processing}

\vspace{-4mm}$$\beta_{m,n}(a) = \displaystyle\min_{\sumover} \left(\sqrt{\sum_{\npIn} \alpha^2_{m,n'}(a_{n'})}\right)$$

\subsection{Min-Max decoding}

\noindent
{\it $\bullet$ Check nodes processing}

\vspace{-4mm}$$\beta_{m,n}(a) = \displaystyle\min_{\sumover} \left(\max_{\npIn} \alpha_{m,n'}(a_{n'})\right)$$

\begin{theo}
Over $\gf(2)$, any decoding algorithm using any one of the $p$-norms, with $p\geq 1$, or the infinity-norm is equivalent to the Min-Sum
decoding. In particular, the Min-Sum, Min-Max, and Euclidian decodings are equivalent.
\end{theo}

\section{Implementation of the Min-Max decoder}
\label{sec:implementations}

In this section we give details about the implementation of the check node processing within the Min-Max decoder. First, we give a {\em
standard implementation} using a well-known forward-backward computation technique. Next, we show that the computations performed within
the standard implementation do not need to use the information concerning all symbols of the Galois field and, consequently, we derive the
so-called {\em selective implementation} of the Min-Max decoder.

\subsection{Standard implementation of the Min-Max decoder}
Let $m$ be a check node and $\N(m) = \{n_1,n_2,\dots, n_d\}$ be the set of variable nodes connected to $m$ in the Tanner graph.

We recursively define forward and backward metrics, $(F_i)_{i=1,d-1}$ and respectively $(B_i)_{i=2,d}$ as follows:

\medskip
\noindent{\bf Forward metrics}

\smallskip\noindent{$\bullet$} 
$F_1(a) = \alpha_{m,n_1}(h_{m,n_1}^{-1}a)$

\smallskip\noindent{$\bullet$} 
    $F_i(a) = \displaystyle \min_{\stackrel{a',a'' \in\gf(q)}{a' + h_{m,n_i}a'' = a}}\left( \max(F_{i-1}(a'),
                 \alpha_{m,n_i}(a''))\right)$

\medskip
\noindent{\bf Backward metrics}

\smallskip\noindent{$\bullet$} 
$B_d(a) = \alpha_{m,n_d}(h_{m,n_d}^{-1}a)$

\smallskip\noindent{$\bullet$} 
    $B_i(a) = \displaystyle \min_{\stackrel{a',a'' \in\gf(q)}{a' + h_{m,n_i}a'' = a}}\left( \max(B_{i+1}(a'),
                 \alpha_{m,n_i}(a''))\right)$

\medskip\noindent
Then check node messages can be computed as follows: 

\smallskip\noindent $\bullet$  $\beta_{m,n_1}(a)=B_2(a)$ and $\beta_{m,n_d}(a) = F_{d-1}(a)$

\smallskip\noindent $\bullet$  $\beta_{m,n_i}(a)= \displaystyle \!\!\!\!\!\min_{\stackrel{a',a'' \in\gf(q)}{a' + a'' =
-h_{m,n_i}a}}\hspace{-3mm}\left( \max(F_{i-1}(a'), B_{i+1}(a''))\right)$

\subsection{Selective implementation of the Min-Max decoder}
In this section we focus on the building blocks of the standard implementation, which are min-max computations of the following type:
$$f(a) = \min_{\stackrel{a',a'' \in\gf(q)}{h'a' + h''a'' = ha}}
     \left( \max(f'(a'), f''(a''))\right)$$
The computation of $f$ requires ${\cal O}(q^2)$ comparisons. Our goal is to reduce this complexity by reducing the number of symbols $a'$
and $a''$ involved in the min-max computation. We start with the following proposition.
\begin{prop}\label{prop_selective}
Let $\Delta', \Delta'' \subset \gf(q)$ be two subsets of the Galois field, such that $\card(\Delta') +
\card(\Delta'') \geq q+1$. Then for any $a\in\gf(q)$, there exist $a'\in\Delta'$ and $a''\in\Delta''$ such
that: $$ha = h'a' + h''a''$$
\end{prop}

\begin{coro}
Let $\Delta', \Delta'' \subset \gf(q)$ be such that the set
   $$\{f'(a')\mid a'\in \Delta'\} \cup \{f''(a'')\mid a''\in \Delta''\}$$
contains the $q+1$ lowest values of the set  
  $$\{f'(a')\mid a'\in \gf(q)\} \cup \{f''(a'')\mid a''\in \gf(q)\}$$
Then for any $a\in\gf(q)$ the following equality holds:
$$ f(a) = \displaystyle\min_{\stackrel{a'\in \Delta',a'' \in\Delta''}{h'a' + h''a'' = ha}} \left( \max(f'(a'), f''(a''))\right)$$
\end{coro}

\begin{exam}
Assume that we have to compute the values of $f$ and the base Galois field is $\gf(8)$. We proceed as follows:

\smallskip\noindent $\bullet$ %
Determine the $9$ smallest values between the $16$ values of the set $\{f'(0),f'(1),\dots, f'(7), f''(0), f''(1),\dots, f''(7)\}$. Let us
assume for instance that the $9$ smallest values are $\{f'(1),f'(2),f'(3),f'(4), f'(6), f'(7), f''(0), f''(1), f''(5)\}$.

\smallskip\noindent $\bullet$ %
Set $\Delta'=\{1, 2, 3, 4, 6, 7\}$ and $\Delta''=\{0, 1, 5\}$, then compute the values of $f$ using only the symbols in $\Delta'$ and
$\Delta''$.

\smallskip\noindent In this way the computation of the $8$ values of $f$ takes only $6\times 3 = 18$ comparisons instead of $64$.
\end{exam}

In general, let $q' = \card (\Delta')$ and $q''=\card(\Delta'')$, where the subsets $\Delta'$ and $\Delta''$ are assumed to satisfy the
conditions of the above corollary (then $q' + q''\geq q+1$). If $q' + q'' = q+1$, then the complexity of the ``min-max'' computation can be
reduced by at least a factor of four, from $q^2$ to $q'q'' \leq \displaystyle\frac{q}{2}(\frac{q}{2}+1)$. The main problem we encounter is
that we should sort the $2q$ values of $f'$ and $f''$ in order to figure out which symbols participate in the sets $\Delta'$ and
$\Delta''$. In order to avoid the sorting process, we generate the subsets:
$$\begin{array}{ccl}
 \Delta'_k & = & \{a'\in\gf(q)\mid \lfloor f'(a') \rfloor = k \}\\
 \Delta''_k & \topspace[5mm]= & \{a''\in\gf(q)\mid \lfloor f''(a'') \rfloor = k \}
\end{array}$$
where $\lfloor x\rfloor$ denotes the integer part of $x$. We then define:
$$\Delta' = \Delta'_0 \cup\cdots\cup\Delta'_t \mbox{ and } \Delta'' = \Delta''_0 \cup\cdots\cup\Delta''_t$$
starting with $t = 0$ and increasing its value until we get $\card(\Delta')+\card(\Delta'') \geq q+1$.

The use of the sets $\Delta'$ and $\Delta''$ has a double interest:
\begin{mylist}
\item it reduces the number of symbols required by (and then the complexity of) the min-max computation,
\item most of the maxima computations become obsolete. Indeed, $\max(f'(a'),
f''(a''))$ is calculated only if symbols $a'$ and $a''$ belong to sets of the same index ({\em i.e.} $a'\in \Delta_{k'}$ and $a''\in
\Delta_{k''}$ with $k' = k''$). Otherwise, the maximum corresponds to the symbol belonging to the set of maximal index.
\end{mylist}

Finally, the proposed approach is carried out in practice as follows:
\begin{mylist}
\item The received a priori information is normalized such that its average value is equal to a predefined constant
$\AI$ (for ``Average a priori Information''). This constant is chosen such that the integer part provides a sharp criterion to distinguish
between the symbols of the Galois field, meaning that sets $\Delta'_k$ and $\Delta''_k$ only contain few elements.

\item We use a predefined threshold $\COT$ (for ``Cut Off Threshold'') representing the maximum
value of variable node messages. Thus, incoming variable node messages greater than or equal to the predefined threshold will not
participate in the min-max computations for check node processing. It follows that the rank $k$ of subsets $\Delta'_k$ and $\Delta''_k$
ranges from $0$ to $\COT-1$. In this case, the subsets $\Delta'_k$ and $\Delta''_k$ may contain all together less than $q+1$ symbols, in
which case the min-max complexity may be significantly reduced. In practice, this generally occurs during the last iterations, when
decoding is quite advanced and doubts persist only about a reduced number of symbols.
\end{mylist}

Constants $\AI$ and $\COT$ have to be determined by Monte Carlo simulation and they only depend on the cardinality of the Galois field.

\section{Simulation results}
\label{sec:simulations}

%

\begin{figure}[!t]
 \includegraphics[width=\linewidth]{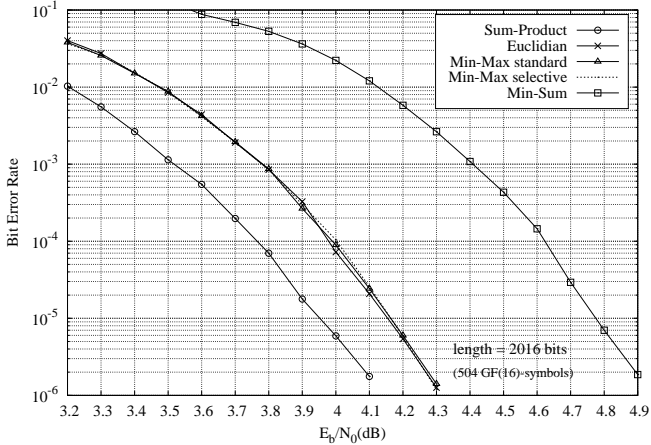}

 \vspace{-5mm}
 \caption{Decoding performance of $\gf(16)$-LDPC codes, AWGN, $16$-QAM, coding rate $1/2$, maximum number of decoding iterations $= 200$}
 \label{fig:courbe_gf4_ber}
\end{figure}

We present Monte-Carlo simulation results for coding rate $1/2$ over the AWGN channel.
Fig. \ref{fig:courbe_gf4_ber} presents the performance of $\gf(16)$-LDPC codes with the $16$-QAM modulation (Galois field symbols are
mapped to constellation symbols). We note that the Euclidian and the Min-Max (standard and selective implementations) algorithms achieve
nearly the same decoding performance, and the gap to the Sum-Product decoding performance is only of  $0.2$~dB
For the selective implementation we used constants $\AI = 12$ and $\COT= 31$.

\begin{figure}[!t]
 \vspace{-4mm}
 \includegraphics[width=\linewidth]{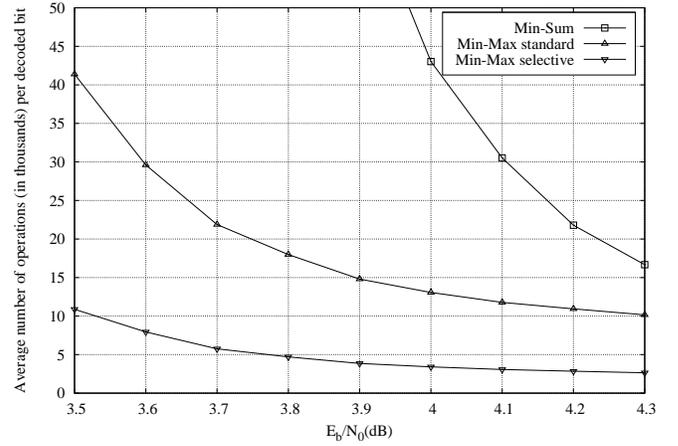}

 \vspace{-5mm}
 \caption{Decoding complexity of $\gf(16)$-LDPC codes, AWGN, $16$-QAM, coding rate $1/2$, maximum number of decoding iterations $= 200$}
 \label{fig:courbe_gf4_cpl}
 \vspace{-5.5mm}
\end{figure}

Fig. \ref{fig:courbe_gf4_cpl} presents the decoding complexity in number of operations per decoded bit (all decoding iterations included)
of the Min-Sum, the standard and the selective Min-Max decoders over $\gf(16)$. The Min-Sum and the standard Min-Max decoders perform
nearly the same number of operations per iteration, but the second decoder has better performance and needs a smaller number of iterations
to converge. On the contrary, the standard and the selective Min-Max decoders have the same performance and they perform the same number of
decoding iterations. Simply, the selective decoder takes a smaller number of operations to perform each decoding iteration, which explains
its lower complexity (by a factor of $4$ with respect to the standard implementation).

\section{Conclusions}
\label{sec:conclusions}

We shown that the extrinsic messages exchanged within the Min-Sum decoding of non binary LDPC codes may be seen as metrics indicating the
distance between a given symbol and the most likely one. By using appropriate metrics, we have derived several low-complexity quasi-optimal
iterative algorithms for decoding non-binary codes. The Euclidian and the Min-Max algorithms are ones of them. Furthermore, we gave a {\em
canonical} selective implementation of the Min-Max decoding, which reduces the number of operations taken to perform each decoding
iteration, without any performance degradation. The quasi-optimal performance together with the low complexity make the Min-Max decoding
very attractive for practical purposes.

\footnotesize
\bibliographystyle{./bib/IEEEbib}
\vspace{-1mm}
\bibliography{./bib/MyBiblio,./bib/Zotero}

\appendices \normalsize
\section{Alternative realizations of the Min-Sum algorithm}
In this section we introduce two alternative realizations, called
$\msazero$ and $\msastar$, of the Min-Sum algorithm. We prove that
they are equivalent to the MSA algorithm of section and, over
$\gf(2)$, the $\msastar$ is equivalent to the Min-Max algorithm. We
assume that the Tanner graph ${\cal H}$ is cycle free thought this
section\footnote{Nevertheless, this assumption can be removed by
replacing ${\cal H}$ with computation trees and ``codewords'' with
``pseudo-codewords''}.

\noindent {\bf Notation}.

\begin{itemize}
\item $n \in {\cal N} = \{ 1, 2, \dots, N \}$ a {\it variable node} of ${\cal H}$.
\item $m \in {\cal M} = \{ 1, 2, \dots, M \}$ a {\it check node} of ${\cal H}$.
\item For each check node $m\in {\cal M}$ and each sequence
$(a_n)_{n\in {\cal H}(m)}$ of $\gf(q)$-symbols indexed by the set of
variable nodes connected to $m$, we note:
$$m\langle (a_n)_{n\in {\cal H}(m)} \rangle = \sum_{n\in {\cal H}(m)} h_{m,n}a_n$$
The sequence $(a_n)_{n\in {\cal H}(m)}$ is said to satisfy the check
node $m$ if $m\langle (a_n)_{n\in {\cal H}(m)} \rangle = 0$. Thus
$$m\langle (a_n)_{n\in {\cal H}(m)} \rangle = 0 \Leftrightarrow
(a_n)_{n\in {\cal H}(m)}\in\L(m)$$
\end{itemize}

\bigskip
The $\msazero$ decoding performs the following computations:

\smallskip\noindent {\bf Initialization} \\
{\it $\bullet$ A priori information}
$$\gamma_n(a) = \ln\left(\frac{{\pr( x_n = 0 \mid \mbox{channel observation} )}}
                               {\pr( x_n = a \mid \mbox{channel observation} )}\right)$$
{\it $\bullet$ Variable to check messages initialization}
$$\alpha_{m,n}(a) = \gamma_n(a)$$

\smallskip\noindent{\bf Iterations}\\
{\it $\bullet$ Check to variable messages}
$$\beta_{m,n}(a) = \displaystyle\min_{\sumover} \left(\sum_{\npIn} \alpha_{m,n'}(a_{n'})\right)$$
{\it $\bullet$ Variable to check messages}
$$\begin{array}{rrl}
\alpha'_{m,n}(a) & = & \displaystyle \gamma_n(a) + \sum_{\mpIn} \beta_{m',n}(a) \\
\alpha_{m,n}(a) & = & \alpha'_{m,n}(a) - \alpha'_{m,n}(0)
\end{array}$$
 {\it $\bullet$ A posteriori information}
$$\tilde{\gamma}_n(a) = \displaystyle \gamma_n(a) + \sum_{\mIn} \beta_{m,n}(a)$$

\begin{theo}
The $\msazero$ decoding converges after finitely many iterations to
$$\tilde{\gamma}_n(a) = \min_{\stackrel{\mathfrak{a}=(a_1,\dots,a_N)}{\mathfrak{a}\in{\cal C}\,:\,a_n = a}}
 \sum_{k\in{\cal N}} \gamma_k(a_k) - \min_{\stackrel{\mathfrak{a}=(a_1,\dots,a_N)}%
 {\mathfrak{a}\in{\cal C}\,:\,\mathfrak{a}\mid_{{\cal V}_n^{(1)}} = 0}}\,\sum_{k\in{\cal N}} \gamma_k(a_k)$$
 where ${\cal V}_n^{(1)}{:}{=} \cup_{m\in{\cal H}(n)} {\cal H}(m)$ is the set of variable nodes
 separated by at most $1$ check node from $n$ (including $n$).
 \end{theo}

 \begin{proof}
 The proof is derived in same manner as that of theorem 3.1 in \cite{Wiberg}.
 The critical point is that we have to control the effect of
 withdrawing the $\alpha_{m,n}(0)$ message from all the others $\alpha_{m,n}(a)$ messages.
 Fix a variable node $n$ and let:
  $$\begin{array}{rcl}
  f(a) & = & \displaystyle \min_{\stackrel{\mathfrak{a}=(a_1,\dots,a_N)}{\mathfrak{a}\in{\cal C}\,:\,a_n = a}}
            \sum_{k\in{\cal N}} \gamma_k(a_k)  \\
  f    & = & \displaystyle\min_{\stackrel{\mathfrak{a}=(a_1,\dots,a_N)}{\mathfrak{a}\in{\cal C}\,:\,\mathfrak{a}
              \mid_{{\cal V}_n^{(1)}} = 0}}\,\sum_{k\in{\cal N}} \gamma_k(a_k) \\
  \tilde{f}(a) & = & f(a) - f
  \end{array}$$
  We have to prove that, after finitely many iterations, the equality $\tilde{\gamma}_n(a)
  =\tilde{f}(a)$ holds. Since the Tanner graph ${\cal H}$ is cycle free, it can be seen as a tree graph rooted at
  $n$. Let $H(n) = \{m_1,m_2,\dots,m_d \}$ and, for $j = 1, 2, \dots, d$, let ${\cal H}_j$ be the sub-graph of
  ${\cal H}$ emanating from the check node $m_j$, as represented below:

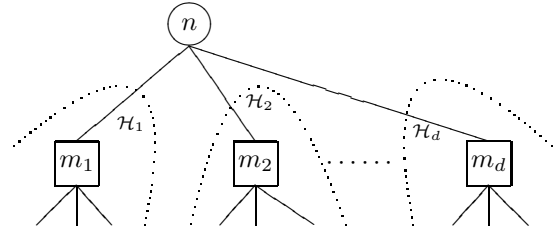
\begin{figure}[h]
  \centering
  $\xymatrix@R=15pt@C=5pt{
  &  &  & *+=[o][F]{\,\,\,n\,\,\,} \ar@{-}'[]!D[ddll]!U \ar@{-}'[]!D[ddr]!U \ar@{-}'[]!D[ddrrrr]!U &  &  &  &  &  \\
  \mbox{ } \ar@{.}@/^3.5pc/[]+<-2mm,-7mm>;[rrdd]!UR_(.7){{\cal H}_1} &   &   &   &   &  &   &   &
  \mbox{ } \ar@{.}@/_3.7pc/[]+<2mm,-7mm>;[lldd]!UL^(.76){{\cal H}_d}                                                                 \\
  & *+=[F]{\,m_1\,}\ar@{-}'[]!D[dl]!UR\ar@{-}'[]!D[d]!U\ar@{-}'[]!D[dr]!UL &                                    &
  & *+=[F]{\,m_2\,}\ar@{-}'[]!D[dl]!UR\ar@{-}'[]!D[d]!U\ar@{-}'[]!D[dr]!UL & \mbox{ }\hspace{1.8mm}\cdots\cdots &
  & *+=[F]{\,m_d\,}\ar@{-}'[]!D[dl]!UR\ar@{-}'[]!D[d]!U\ar@{-}'[]!D[dr]!UL &                                      \\
  &  & \mbox{ }\hspace{1mm} & \mbox{ }\hspace{5mm}\ar@{.}@/^4.5pc/[rr]!U_{{\cal H}_2}
  &  & \mbox{ }\hspace{5mm} & \mbox{ }\hspace{2mm} &  &                                                           }$
  \caption{Graphical representation of sub-graphs ${\cal H}_j$} \label{graphs_Hj}
\end{figure}

\noindent Let also ${\cal C}_{j}$ be the linear code corresponding
to the sub-graph ${\cal H}_j \cup \{ n
  \}$. Since $\gamma_k(0) = 0$ for any $k=1,\dots,N$, we have $\displaystyle f = \sum_{j=1}^{d} f_j$,
  where $f_j = \displaystyle\min_{\stackrel{\mathfrak{a}\in{\cal C}_{j}}{\mathfrak{a}
               \mid_{H(m_j)} = 0}}\,\sum_{k\in{\cal H}_{j}} \gamma_k(a_k)$. Therefore,
$$\begin{array}{rcl}
 \tilde{f}(a) & = & \displaystyle \left(\min_{\stackrel{\mathfrak{a}=(a_1,\dots,a_N)}{\mathfrak{a}\in{\cal C}\,:\,a_n = a}}
            \sum_{k\in{\cal N}} \gamma_k(a_k)\right) - f \\
            & = & \displaystyle \gamma_n(a) + \sum_{j=1}^{d} \left(\min_{\stackrel{\mathfrak{a}
            \in{\cal C}_{j}}{a_n = a}}\sum_{k\in{\cal H}_j}\gamma_k(a_k)\right) - \sum_{j=1}^{d} f_j
 \end{array}$$
 Denoting $g_{m_j}(a) = \displaystyle \min_{\stackrel{\mathfrak{a}
            \in{\cal C}_{j}}{a_n = a}}\left(\sum_{k\in{\cal H}_j}\gamma_k(a_k)\right) - f_j $ we get
 $$\tilde{f}(a)  =\displaystyle \gamma_n(a) + \sum_{j=1}^{d}g_{m_j}(a)$$
 This formula has the same structure as the update rule of the a posteriori information
 $\tilde{\gamma}_n(a)$. It follows that the equality $\tilde{\gamma}_n(a)
  =\tilde{f}(a)$ holds provided that $\beta_{m_j,n}(a) = g_{m_j}(a)$. Due to the symmetry of the situation it
  suffices to carry out the case $j=1$. For this, let $H(m_1) = \{n,n_1,n_2,\dots,n_d \}$ and,
  for $j = 1, 2, \dots, d$, let ${\cal H}_{1,j}$ be the sub-graph of ${\cal H}_1$ emanating from
the variable node $n_j$, as represented below:

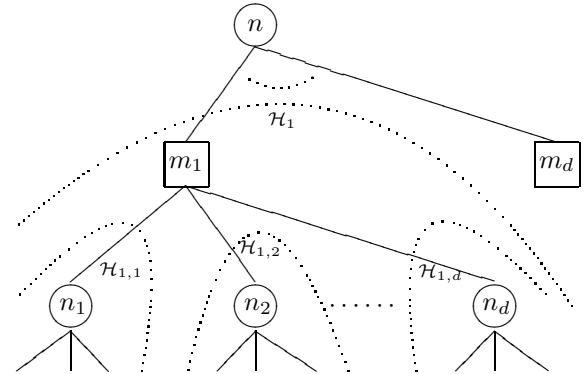
\begin{figure}[h]
  \centering
  $\xymatrix@R=15pt@C=5pt{
  &  &  &  & *+=[o][F]{\,\,\,n\,\,\,} \ar@{-}'[]!D[ddl]!U^(.25)\cdot="a"\ar@{-}'[]!D[ddrrrr]!U_(.2)\cdot="b" &
  &  &  & \ar@{.}@/_0.3pc/ "a"+0;"b"+0 \\
  \mbox{ }  &   &   &   &   &  &   &   &                        \\
  &  &  &  *+=[F]{\,m_1\,}\ar@{-}'[]!D[ddll]!U \ar@{-}'[]!D[ddr]!U \ar@{-}'[]!D[ddrrrr]!U &  &  &  &  & *+=[F]{\,m_d\,} \\
  \mbox{ }\ar@{.}@/^5.2pc/[drrrrrrrr]+<2mm,0mm>_(.45){{\cal H}_1}  &   &   &   &   &  &   &   &                        \\
  \mbox{ }\hspace{3mm}\ar@{.}@/^4pc/[rrd]!UR_(.75){{\cal H}_{1,1}}& *+=[o][F]{\,\,\,n_1\,\,}\ar@{-}'[]!D[dl]!UR\ar@{-}'[]!D[d]!U\ar@{-}'[]!D[dr]!UL &                           &
  & *+=[o][F]{\,\,\,n_2\,\,}\ar@{-}'[]!D[dl]!UR\ar@{-}'[]!D[d]!U\ar@{-}'[]!D[dr]!UL & \mbox{ }\hspace{1.8mm}\cdots\cdots&
  & *+=[o][F]{\,\,\,n_d\,\,}\ar@{-}'[]!D[dl]!UR\ar@{-}'[]!D[d]!U\ar@{-}'[]!D[dr]!UL & \mbox{ }\hspace{3mm}\ar@{.}@/_4pc/[lld]!UL^(.76){{\cal H}_{1,d}}  \\
  &  & \mbox{ }\hspace{1mm} & \mbox{ }\hspace{5mm}\ar@{.}@/^4.5pc/[rr]!U_{{\cal H}_{1,2}}
  &  & \mbox{ }\hspace{5mm} & \mbox{ }\hspace{2mm} &  &                                                           }$
  \caption{Graphical representation of subgraphs ${\cal H}_{1,j}$} \label{graphs_H1j}
\end{figure}
\noindent Let also ${\cal C}_{1,j}$ be the linear code corresponding
to the sub-graph ${\cal H}_{1,j}$ and define $\displaystyle t_j(a_j)
=\min_{\stackrel{\mathfrak{a}\in{\cal C}_{1,j}}{a_{n_j}=a_j}}
\left(\sum_{k\in{\cal H}_{1,j}}\gamma_k(a_k)\right)$. Then:
$$\min_{\stackrel{\mathfrak{a} \in{\cal C}_{1}}{a_n = a}}\left(\sum_{k\in{\cal H}_1}\gamma_k(a_k)\right) =
 \min_{\stackrel{(a_1,\dots,a_d)\in \mbox{\rm\sc\scriptsize gf}(q)^d}{m_1\langle a,a_1,\dots,a_d\rangle = 0}}
 \left(\sum_{j=1}^{d} t_j(a_j)\right)$$
 and
 $$\begin{array}{r@{\ }c@{\ }l}
 f_1 & = &\displaystyle\min_{\stackrel{\mathfrak{a}\in{\cal C}_{1}}{\mathfrak{a}
               \mid_{H(m_1)} = 0}}\,\sum_{k\in{\cal H}_{1}} \gamma_k(a_k) =
               \sum_{j=1}^{d} \min_{\stackrel{a\in {\cal C}_{1,j}}{a_{n_j}=0}}\left(\sum_{k\in{\cal H}_{1,j}}
               \gamma_k(a_k)\right) \\
     & = & \displaystyle\sum_{j=1}^{d}t_j(0)
 \end{array}$$
 Therefore:
 $$\begin{array}{rcl}
 g_{m_1}(a) & = & \displaystyle \min_{\stackrel{(a_1,\dots,a_d)\in \mbox{\rm\sc\scriptsize gf}(q)^d}
                   {m_1\langle a,a_1,\dots,a_d\rangle = 0}} \left(\sum_{j=1}^{d} t_j(a_j)\right) -
                   \sum_{j=1}^{d}t_j(0)\\
            & = & \displaystyle \min_{\stackrel{(a_1,\dots,a_d)\in \mbox{\rm\sc\scriptsize gf}(q)^d}
                   {m_1\langle a,a_1,\dots,a_d\rangle = 0}} \left(\sum_{j=1}^{d} \left(t_j(a_j) - t_j(0)\right)\right)
 \end{array}$$
Defining $t = \displaystyle\min_{\stackrel{\mathfrak{a}\in{\cal
C}_{1,j}}{\mathfrak{a}\mid_{{\cal V}_n^{(1)}} = 0}} \left(
\sum_{k\in{\cal H}_{1,j}} \gamma_k(a_k)\right)$,
$$\begin{array}{r@{\ }c@{\ }l}
 f'_{n_j}(a_j) & = & \displaystyle\min_{\stackrel{\mathfrak{a}\in{\cal C}_{1,j}}{a_{n_j}=a_j}}
           \,\,\sum_{k\in{\cal H}_{1,j}}\gamma_k(a_k) - \min_{\stackrel{\mathfrak{a}\in{\cal C}_{1,j}}
           {\mathfrak{a}\mid_{{\cal V}_n^{(1)}} = 0}} \,\,\sum_{k\in{\cal H}_{1,j}} \gamma_k(a_k)  \\
   & = & t_j(a_j) - t
\end{array}$$
and $f_{n_j}(a_j) = f'_{n_j}(a_j) - f'_{n_j}(0) = t_j(a_j) - t_j(0)$
we get
$$ g_{m_1}(a) =  \min_{\stackrel{(a_1,\dots,a_d)\in \mbox{\rm\sc\scriptsize gf}(q)^d}
                   {m_1\langle a,a_1,\dots,a_d\rangle = 0}} \left(\sum_{j=1}^{d} f_{n_j}(a_j) \right)$$
This formula has the same structure as the update rule of the check
to variable messages
 $\beta_{m_1,n}(a)$. It follows that the equality $\beta_{m_1,n}(a) = g_{m_1}(a)$ holds, provided that
$\alpha_{m_1,n_j}(a) = f_{n_j}(a)$. This derives from the fact that
$f_{n_j}(a)$ verifies the same update rule as $\alpha_{m_1,n_j}(a)$
or, equivalently, $f'_{n_j}(a)$ verifies the same update rule as
$\alpha'_{m_1,n_j}(a)$. The proof can be proceeded in same manner as
for $f(a)$, and then will be omitted.

This process is repeated recursively until the leaf variable nodes
are reached. Finally, we remark that if $n_j$ where a leaf variable
node, then $f_{n_j}(a_j) = \gamma_{n_j}(a_j) =
\alpha_{m_1,n_j}(a_j)$ and so we are done.
\end{proof}

\bigskip
The $\msastar$ decoding performs the following computations:

\smallskip\noindent {\bf Initialization} \\
{\it $\bullet$ A priori information}
$$\gamma_n(a) = \ln\left(\frac{{\pr( x_n = s_n \mid \mbox{channel observation} )}}
                               {\pr( x_n = a \mid \mbox{channel observation} )}\right)$$
where $s_n$ is the most likely symbol for $x_n$.

\smallskip\noindent {\it $\bullet$ Variable to check messages initialization}
$$\alpha_{m,n}(a) = \gamma_n(a)$$

\smallskip\noindent{\bf Iterations}\\
{\it $\bullet$ Check to variable messages}
$$\beta_{m,n}(a) = \displaystyle\min_{\sumover} \left(\sum_{\npIn} \alpha_{m,n'}(a_{n'})\right)$$
{\it $\bullet$ Variable to check messages}
$$\begin{array}{rrl}
\alpha'_{m,n}(a) & = & \displaystyle \gamma_n(a) + \sum_{\mpIn} \beta_{m',n}(a) \\
\alpha'_{m,n}    &   & \displaystyle \min_{a\in\gf(q)}\alpha'_{m,n}(a) \\
 \alpha_{m,n}(a) & = & \alpha'_{m,n}(a) - \alpha'_{m,n}
\end{array}$$
 {\it $\bullet$ A posteriori information}
$$\tilde{\gamma}_n(a) = \displaystyle \gamma_n(a) + \sum_{\mIn} \beta_{m,n}(a)$$

Fixe a variable node $n$, note ${\cal H}_{\star} = {\cal
H}\setminus\left(\{n\}\cup {\cal H}(n)\right)$ and let ${\cal
C}_{\star}$ be the linear code associated with ${\cal H}_{\star}$
(when a node of ${\cal H}$ is removed we also remove all the edges
that are incident to the node). With this notation we have the
following theorem.

\begin{theo}
The $\msastar$ decoding converges after finitely many iterations to
$$\tilde{\gamma}_n(a) = \min_{\stackrel{\mathfrak{a}=(a_k)_{k\in{\cal N}}}{\mathfrak{a}\in{\cal C}\,:\,a_n = a}}
 \,\sum_{k\in{\cal N}} \gamma_k(a_k) - \min_{\stackrel{\mathfrak{a}=(a_k)_{k\in{\cal H}_\star}}%
 {\mathfrak{a}\in{\cal C}_\star}}\,\sum_{k\in{\cal H}_\star} \gamma_k(a_k)$$
 \end{theo}
The proof can be derived in same manner os that of the above
theorem, and then will be omitted.

\bigskip We also note that the convergence formula of the MSA decoder
 is (see theorem 3.1 in \cite{Wiberg}):
$$\tilde{\gamma}_n(a) = \min_{\stackrel{\mathfrak{a}=(a_k)_{k\in{\cal N}}}{\mathfrak{a}\in{\cal C}\,:\,a_n = a}}
 \,\sum_{k\in{\cal N}} \gamma_k(a_k)$$
 For each of the $\msazero$ and $\msastar$ decoders the convergence formula is obtained by withdrawing from the right hand side term\footnote{Note also that
the a priori informations $\gamma_n(a)$ are not computed in the same
way for the three decoders but they only differ by a term
independent of $a$. Thus, for the MSA decoder $\gamma_n(a) =
-\ln(\pr(x_n = a))$, for the $\msazero$ decoder $\gamma_n(a) =
-\ln(\pr(x_n = a)) + \ln(\pr(x_n = 0))$ and for the $\msastar$
decoder $\gamma_n(a) = -\ln(\pr(x_n = a)) + \ln(\pr(x_n = s_n))$,
all probabilities being conditioned by the channel observation.} of
the above formula a term that is independent of $a$. Note also that
similar formulas hold after any fixed number of iterations - to see
this, it suffices to cut off the tree graph ${\cal H}$ at the
appropriate depth. Therefore, we get the following:
\begin{coro}
The MSA, $\msazero$ and $\msastar$ decoders are equivalent.
\end{coro}

\bigskip
We prove now that the MSA and the MMA decoders are equivalent over
$\gf(2)$ (in a similar maner can be proved that over $\gf(2)$ the
MSA decoder is equivalent to any other Min-$\parallel\,\parallel_p$
decoder.)

\begin{proof}
Since MSA and $\msastar$ decoder are equivalent, it suffices to
prove that $\msastar$ and MMA decoders are equivalent over $\gf(2)$.
In this case, the variable to check messages $\alpha_{m,n}(a)$ are
non negative and they concern only two symbols $a\in\{0,1\}$.
Moreover, for one of these symbols the corresponding variable to
check message in zero (precisely, for the symbol realizing the
minimum of $\{\alpha'_{m,n}(a), a=0,1\}$). Fix a check node $m$, a
variable node $n$ and a symbol $a\in \gf(2)$, and let
$(\bar{a}_{n'})_{n'\in{\cal H}(m)\setminus\{n\}}$, such that
$m\langle (\bar{a}_{n'})_{n'},a \rangle = 0$, be the sequence
realizing the minimum:
$$\displaystyle\min_{\sumover} \left(\sum_{\npIn} \alpha_{m,n'}(a_{n'})\right)$$
Then, for the $\msastar$ decoder
$$\displaystyle\beta_{m,n}(a) =\sum_{\npIn} \alpha_{m,n'}(\bar{a}_{n'})$$
Suppose that there are two symbols, say $\bar{a}_{n'_1}$ and
$\bar{a}_{n'_2}$, of the sequence $(\bar{a}_{n'})_{n'\in{\cal
N}(m)\setminus\{n\}}$, whose corresponding variable to check
messages are not zero, i.e. $\alpha_{m,n'_i}(\bar{a}_{n'_i}) > 0,
i=1,2$. Then, consider a new sequence
$(\bar{\bar{a}}_{n'})_{n'\in{\cal H}(m)\setminus\{n\}}$, such that
$\bar{\bar{a}}_{n'} = {\bar{a}}_{n'}$ if $n' \not = n'_i$ and
$\bar{\bar{a}}_{n'_i} = {\bar{a}}_{n'_i}+1\mod 2$. This sequence
still satisfies the check $m$, i.e. $m\langle
(\bar{\bar{a}}_{n'})_{n'},a \rangle = 0$ and
$$\sum_{\npIn} \alpha_{m,n'}(\bar{\bar{a}}_{n'}) < \sum_{\npIn} \alpha_{m,n'}(\bar{a}_{n'})$$
as $\alpha_{m,n'}(\bar{\bar{a}}_{n'_i}) = 0 <
\alpha_{m,n'}(\bar{a}_{n'_i})$, what contradicts the minimality of
the sequence $(\bar{a}_{n'})_{n'\in{\cal H}(m)\setminus\{n\}}$. It
follows that there is at mot one symbol $\bar{a}_{n'}$ whose
corresponding variable to check message
$\alpha_{m,n'}(\bar{a}_{n'})\not = 0$. Consequently, the sequence
$(\bar{a}_{n'})_{n'\in{\cal H}(m)\setminus\{n\}}$ also realizes the
minimum
$$ \min_{\sumover} \left(\max_{\npIn} \alpha_{m,n'}(a_{n'})\right)$$
so the update rules for check to variable messages computes the same
messages for both decoders.
\end{proof}

\section{Proof of Proposition \ref{prop_selective}}

\begin{proof}
Let $a\in \gf(q)$. Consider the functions $\varphi,
\psi:\gf(q)\rightarrow\gf(q)$, defined by $\varphi(x) = ha+h'x$, and
$\psi(x) = h''x$. Since $\varphi$, and $\psi$ are injective, we
have:
$$\card(\varphi(\Delta')) + \card(\psi(\Delta'')) = \card(\Delta') + \card(\Delta'') \geq q+1$$
It follows that $\varphi(\Delta')\cap\psi(\Delta'')\not= \emptyset$,
so there are $a'\in\Delta'$, and $a''\in\Delta''$ such that
$$\varphi(a') = \psi(a'') \Leftrightarrow ha+h'a' = h''a'' \Leftrightarrow a =
h^{-1}(h'a' + h''a'')$$
\end{proof}

\end{document}